# HYDRODYNAMIC SIMULATIONS OF GALAXY FORMATION.

## II. PHOTOIONIZATION AND THE FORMATION OF LOW MASS GALAXIES.

Anne A. Thoul[1,2] and David H. Weinberg[1,3]

E-mail: athoul@cfa.harvard.edu, dhw@payne.mps.ohio-state.edu

## ABSTRACT

Photoionization by the high-redshift ultraviolet radiation background heats low density gas before it falls into dark matter potential wells, and it eliminates the neutral hydrogen and singly ionized helium that dominate cooling of primordial gas at temperatures of $10^4 - 10^5$K. We investigate the influence of photoionization on galaxy formation using high-resolution simulations with a 1-dimensional, spherically symmetric, Lagrangian hydrodynamics/gravity code. We find that the presence of a photoionizing background suppresses the formation of galaxies with circular velocities $v_{circ} \lesssim 30 \, \mathrm{km \, s^{-1}}$ and substantially reduces the mass of cooled baryons in systems with circular velocities up to $v_{circ} \sim 50 \, \mathrm{km \, s^{-1}}$. Above $v_{circ} \sim 75 \, \mathrm{km \, s^{-1}}$, photoionization has no significant effect. Photoionization exerts its influence primarily by heating gas before collapse; the elimination of line cooling processes is less important. We discuss the implications of these results for hierarchical theories of galaxy formation.

*Subject headings:* Galaxies: formation, Hydrodynamics, Methods: numerical

## 1. Introduction

Contemporary cosmological theories imply that the galaxies and the large-scale structure observed today have evolved by gravitational instability from small-amplitude primordial density fluctuations in the early universe. In theories where the universe is dominated by cold, collisionless dark matter, there is no natural cutoff in the primordial power spectrum at small scales. The clustering of dark matter in such theories is hierarchical: low mass objects collapse first, then merge into progressively larger systems. However, the material that we observe directly in galaxies is baryonic, so its evolution is influenced by hydrodynamic and thermodynamic processes such as shocks, radiative cooling, and photoionization, and these processes can introduce natural scales into the baryon distribution. In particular, pressure gradients can suppress gravitational collapse on scales below some characteristic mass, usually referred to as the Jeans mass.

---

[1] Institute for Advanced Study, Princeton, NJ 08540

[2] Harvard-Smithsonian Center for Astrophysics, Cambridge, MA 02138

[3] Ohio State University, Department of Astronomy, Columbus, OH 43210

In this paper we study galaxy formation in the presence of photoionization by an ultraviolet (UV) radiation background, using simulations of spherically symmetric collapses of mixed baryon/dark matter perturbations. The potential importance of photoionization was emphasized by Efstathiou (1992; see also Shapiro, Giroux & Babul 1994), who suggested it as a mechanism to suppress the formation of dwarf galaxies in hierarchical models. According to semi-analytic calculations, hierarchical models naturally predict a faint-end slope of the galaxy luminosity function that is much steeper than most observational estimates (e.g. White & Frenk 1991; Kauffman, White & Guideroni 1993; Cole et al. 1994). Various solutions to this problem have been proposed, including feedback from star formation (Dekel and Silk 1986), reheating of the intergalactic medium by supernova winds (Tegmark, Silk, & Evrard 1993), Compton heating from energetic objects at very high redshifts (Blanchard, Valls-Gabaud, & Mamon 1992), and rapid merging of dwarf galaxies between their formation and $z = 0$. However, photoionization would be an attractively conservative answer to the faint galaxy conundrum because the spectra of quasars provide direct evidence that the intergalactic medium was highly photoionized out to redshifts $z \sim 5$ (e.g., Schneider, Schmidt, & Gunn 1989; Webb et al. 1992; Meiksin & Madau 1993). Even though some gas must have collapsed and cooled early to produce stars and/or AGNs, the sources of the background radiation, it is plausible that most of the gas was still relatively homogeneous at the time that it was reionized. Photoionization can have two distinct effects on pre-galactic or proto-galactic gas: it heats gas which would otherwise be at very low temperature, and it reduces cooling rates by lowering the fraction of neutral atoms. As Efstathiou (1992) pointed out, both effects can influence the ability of gas to cool and condense in low mass systems. However, because of the simplified nature of Efstathiou's analytic model, his estimate of the quantitative impact of photoionization was very uncertain.

Thoul and Weinberg (1995; hereafter Paper I) developed a 1-dimensional, spherically symmetric, Lagrangian hydrodynamics/gravity code to study the effects of radiative cooling on the high-mass end of the luminosity function. In this paper, we use this code to study the effects of radiative cooling and photoionization on the formation of low mass systems. While 1-d simulations require a geometrical idealization, they complement more computationally intensive calculations with 3-d N-body/hydro codes (e.g. Katz, Hernquist & Weinberg 1992; Cen & Ostriker 1993; Evrard, Summers & Davis 1994; Navarro & White 1994; Steinmetz & Muller 1994) in several important ways. First, 1-d simulations can achieve much higher spatial resolution and mass resolution, so they compute hydrodynamic and thermodynamic effects more accurately. Second, with faster simulations one can investigate a broad parameter space, and one can gain physical insight by examining how variations of inputs change the final results. Finally, 1-d simulations are easier to visualize and interpret than their 3-d cousins, again making it easier to gain physical understanding of their results.

In §2 below, we describe our treatment of photoionization and radiative cooling and discuss the initial conditions used for the simulations in this paper. Section 3 presents the results of the simulations. We discuss their implications in §4.

## 2. Model

Our simulations evolve a spherically symmetric density enhancement in an Einstein-de Sitter ($\Omega = 1$) universe. We assume that collisionless dark matter and baryons are initially well mixed, with a baryon fraction $\Omega_b = 0.1$. The collisionless component is described by the equation of motion $dv_d/dt = -GM(r_d)/r_d^2$, where $r_d$ and $v_d$ are the radii and velocities of the collisionless mass shells and $M(r_d)$ is the total (baryonic and dark matter) mass inside radius $r_d$. The baryon component is described by the fluid equations for a perfect gas moving in the gravitational field of both components. We use the standard, second-order accurate, Lagrangian finite-difference scheme (Bowers & Wilson 1991). Paper I describes the code in detail and presents many tests on problems with analytically known solutions.

### 2.1. Cooling and heating

The radiative cooling is calculated for a gas of primordial composition, 76% hydrogen and 24% helium by mass. We compute the abundances of ionic species as a function of density and temperature by assuming that the gas is in ionization equilibrium with a spatially uniform background of UV radiation. In other words, we choose the abundances so that the rate at which each species is depopulated by photoionization, collisional ionization, or recombination to a less ionized state is equal to the rate at which it is populated by recombination from a more ionized state or by photoionization or collisional ionization of a less ionized state. Ionization equilibrium should be an excellent approximation under the physical conditions that are relevant for these simulations (see, for example, the discussion of Vedel et al. 1994). The full set of equations that we use to obtain abundances and cooling rates is listed in Katz, Weinberg & Hernquist (1995). The original source for most of these formulae is Black (1981), and we adopt the high-temperature corrections of Cen (1992).

We model the UV background spectrum as a power-law with a redshift-dependent normalization,

$$J_\gamma(\nu, z) = J_0(z) \left(\frac{\nu_L}{\nu}\right)^\alpha,$$ (1)

where $\nu_L$ is the Lyman limit frequency ($h\nu_L = 13.6$ eV), $\alpha$ is the spectral slope, and

$$J_0(z) \equiv J_{-21}(z) \times 10^{-21} \text{ erg s}^{-1} \text{ cm}^{-2} \text{ sr}^{-1} \text{ Hz}^{-1}$$ (2)

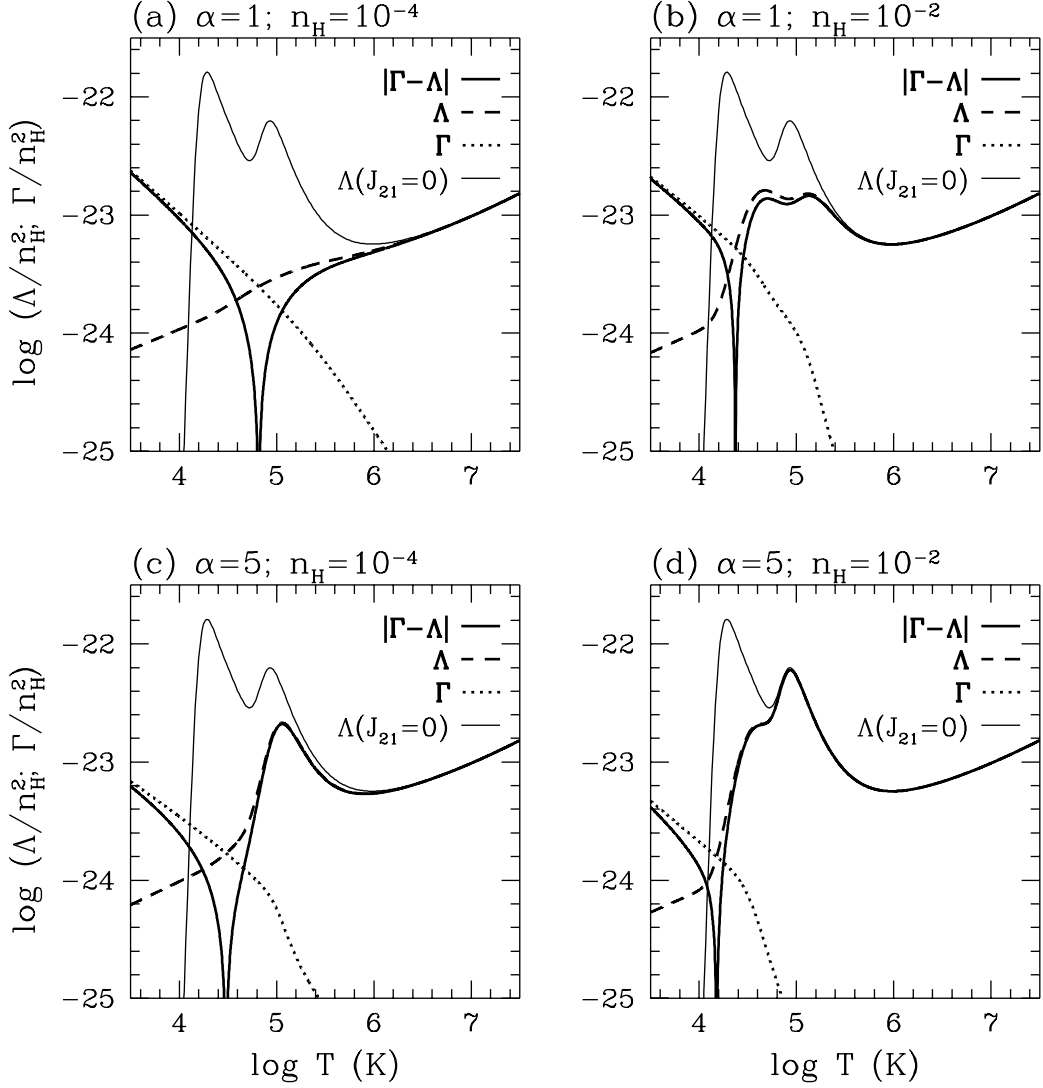

Fig. 1.— Cooling and heating rates as a function of temperature for a gas of primordial composition ($f_H = n_H m_p/\rho_g = 0.76$, $f_{He} = 0.24$) and a UV background intensity $J_{-21} = 1$ (thick lines) and $J_{-21} = 0$ (thin lines). The thick dashed lines are the cooling rates, the thick dotted lines are the heating rates, and the thick solid lines are the net ($|\Gamma - \Lambda|$) cooling/heating rates, all scaled by $n_H^{-2}$. Abundances of ionic species are computed assuming ionization equilibrium. The units of $\Lambda$ and $\Gamma$ are ergs s$^{-1}$ cm$^{-3}$.

is the amplitude. Predictions of the UV background from QSOs and observations of the proximity effect in the Ly$\alpha$ forest suggest that $J_{-21} \sim 0.3 - 1$ at $z \sim 2 - 4$. In Figure 1 we show the cooling and heating rates $\Gamma$ and $\Lambda$ and the net cooling/heating rate $|\Gamma - \Lambda|$ as a function of temperature. In the presence of an ionizing background, the cooling curves are also function of density and of the intensity and slope of the UV spectrum. In panels (a) and (b) we show the cooling curves for $n_H = 10^{-4}\,\mathrm{cm}^{-3}$ and $n_H = 10^{-2}\,\mathrm{cm}^{-3}$, $\alpha = 1$, and $J_{-21} = 1$. In panels (c) and (d), we show the cooling curves for the same values of $n_H$ and $J_{-21}$ but with $\alpha = 5$. Here $n_H$ is the hydrogen number density, which can be conveniently expressed in the form

$$n_H = 2.31 \times 10^{-5} \left(\frac{1+z}{3}\right)^3 \left(\frac{\Omega_b h^2}{0.1}\right) \frac{\rho}{\rho_H}\,\mathrm{cm}^{-3}, \tag{3}$$

where $z$ is the redshift, $\rho/\rho_H$ is the ratio of the local density to the critical density $\rho_H \equiv (6\pi G t^2)^{-1}$, $\Omega_b$ is the baryon density parameter, and $h \equiv H_0/(100\,\mathrm{km\,s}^{-1}\,\mathrm{Mpc}^{-1})$. For comparison, Figure 1 also shows the cooling curve obtained in the absence of a UV background.

As Figure 1 illustrates, the UV radiation affects the thermal state of protogalactic gas in two ways. The residual energy of photoelectrons heats low-temperature gas, and photoionization eliminates the neutral hydrogen and singly ionized helium that dominate cooling at temperatures $T \sim 10^4 - 10^5\,\mathrm{K}$. The zero-crossings in Figure 1 indicate the equilibrium temperature $T_{eq}$, where cooling and heating balance. Values of $T_{eq}$ lie in the range $10^4 - 10^5\,\mathrm{K}$, depending on the gas density and on the UV background spectrum. A harder spectrum yields more energetic photoelectrons and a higher $T_{eq}$. Conversely, raising the gas density with a fixed UV background increases cooling and lowers $T_{eq}$. In the low density limit ($n_H \lesssim 10^{-6}\,\mathrm{cm}^{-3}$ for $J_{-21} \sim 1$), $T_{eq}$ becomes independent of density because a drop in the recombination cooling rate is compensated by a drop in the number of neutral atoms able to absorb energy by being photoionized.

From Figure 1, we can already begin to guess how the impact of photoionization on a collapsing perturbation might depend on the parameters of the perturbation and of the UV background. Above $T \sim 10^6\,\mathrm{K}$, photoionization does not influence cooling curves at all, so we expect no impact on the formation of objects with virial temperature above this value (corresponding to a circular velocity $v_{circ} \approx 160\,\mathrm{km\,s}^{-1}$, see equation [13] below). For a softer background spectrum, e.g. $\alpha = 5$ instead of $\alpha = 1$, the cooling bump at $10^5\,\mathrm{K}$ from singly ionized helium returns at moderate density, so the temperature above which cooling rates are minimally affected is lower. Furthermore, if (as we shall eventually argue) the heating of low temperature gas is more important than the reduction of cooling rates, then the relevant temperature to consider is $T_{eq}$, which is considerably lower than the temperature at which cooling rates are unaffected. Again one expects a harder spectrum to have a greater impact. Finally, since equilibrium temperatures and reductions in cooling rates are smaller when the gas density is high, we expect photoionization to have less effect on objects that collapse at higher redshifts.

## 2.2. Perturbations

As initial conditions, we adopt the average density profile around a $2\sigma$ peak in a Gaussian random density field of power spectrum $Ak^{-2}e^{-k^2R_f^2}$ (Bardeen et al. 1986), as shown in Figure 2. The $n = -2$ power-law index is roughly the effective index of the cold dark matter power spectrum on galaxy mass scales. These initial conditions have two free parameters, $A$ and $R_f$, which we can specify more physically by the filter mass $M_f$, defined as the mass contained within a sphere of radius $2R_f$ where $R_f$ is the filter radius, and the collapse redshift $z_c$, defined to be the redshift at which the $r = 2R_f$ shell would collapse to $r = 0$ in the absence of pressure. We cast our definitions in terms of $2R_f$ instead of $R_f$ because it is the radius at which the perturbation profile has fallen to roughly half of its peak value.

We will often characterize a perturbation by the circular velocity

$$v_{circ} = \sqrt{\frac{GM_f}{0.5r_{ta}}} \tag{4}$$

that would be associated with the shell surrounding mass $M_f$ if it collapsed and virialized at half its turnaround radius $r_{ta}$. The circular velocity can be expressed in terms of $M_f$ and the collapse redshift $z_c$. The turnaround radius is given by

$$r_{ta} = r_i \left(\frac{5}{3}\overline{\delta}_i\right)^{-1}, \tag{5}$$

where $\overline{\delta}_i$ and $r_i$ are the mean interior density contrast and the radius at some redshift $z_i$ high enough that the perturbation is still in the linear regime. The collapse time $t_c$ is given by

$$t_c = t_i \left(\frac{3\pi}{2}\right)\left(\frac{5}{3}\overline{\delta}_i\right)^{-3/2}, \tag{6}$$

where $t_i$ is the time corresponding to the redshift $z_i$, and the mass $M_f$ is given by

$$M_f = \frac{4}{3}\pi r_i^3 \rho_c (1+z_i)^3, \tag{7}$$

where $\rho_c$ is the critical density of the universe today. From equations (5)-(7), and using the relation

$$\frac{t_c}{t_i} = \left(\frac{1+z_i}{1+z_c}\right)^{3/2} \tag{8}$$

we get

$$r_{ta} = \left(\frac{M_f}{3\pi^3 \rho_c}\right)^{1/3} (1+z_c)^{-1}. \tag{9}$$

The circular velocity can therefore be written as

$$v_{circ} = \left[2GM_f^{2/3}\left(3\pi^3\rho_c\right)^{1/3}(1+z_c)\right]^{1/2} = 128\,\mathrm{km\,s^{-1}} \left(\frac{M_f}{10^{11}h^{-1}M_\odot}\right)^{1/3}\left(\frac{1+z_c}{3}\right)^{1/2}. \tag{10}$$

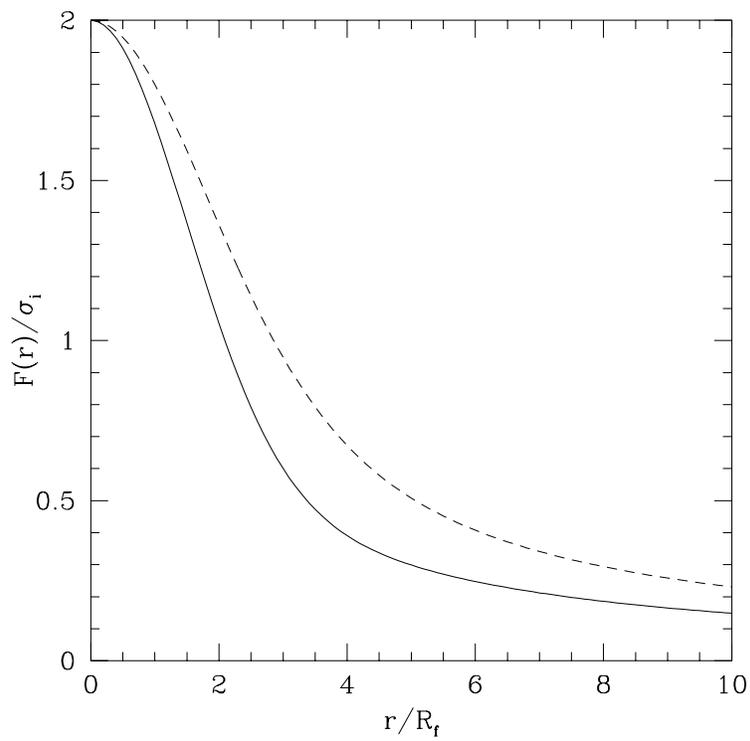

Fig. 2.— Mean profile of a $2\sigma$ peak in a Gaussian density field with power spectrum $P(k) = Ak^{-2}e^{-k^2R_f^2}$. The solid line shows the density contrast; the dashed line shows the mean interior density contrast. We adopt this profile as the initial density distribution for our collapse simulations.

It is natural to express results in terms of a circular velocity rather than a perturbation mass because we can directly associate a temperature with this circular velocity. Let us assume that the local density profile is $\rho(r) = \rho_1(r/r_1)^{-2}$ and demand that the gas be isothermal and in hydrostatic equilibrium. In this case, the gravitational force on a shell of width $dr$ enclosing a mass $M$ is balanced by the differential pressure force,

$$\frac{GM}{r^2} 4\pi r^2 \rho(r)dr = \left(-\frac{dp}{dr}dr\right) 4\pi r^2, \tag{11}$$

where the pressure $p$ is given by

$$p = \frac{\rho k_B T}{\mu m_p}. \tag{12}$$

Here, $k_B$ is the Boltzman constant, $p$, $T$ and $\mu$ are the pressure, temperature, and molecular weight of the gas, and $m_p$ is the proton mass. With $v_{circ} = \sqrt{GM/r}$, equations (11) and (12) yield

$$T = \left(\frac{\mu m_p}{2k_B}\right) v_{circ}^2 = 5.95 \times 10^5 \,\mathrm{K} \left(\frac{\mu}{0.6}\right) \left(\frac{v_{circ}}{128\,\mathrm{km\,s^{-1}}}\right)^2. \tag{13}$$

This result can also be obtained by equating the kinetic energy of a mass $M$ moving at the rms velocity $v_{rms}$ to the internal energy of a mass $M$ at temperature $T$,

$$\frac{1}{2}Mv_{rms}^2 = \frac{3}{2}k_B T \left(\frac{M}{\mu m_p}\right). \tag{14}$$

For a singular isothermal sphere, the rms velocity is related to the circular velocity by $v_{rms} = \sqrt{3/2}\,v_{circ}$ (Binney & Tremaine 1987). Substituting into equation (14), we get

$$\frac{1}{2}Mv_{circ}^2 = k_B T \left(\frac{M}{\mu m_p}\right) \tag{15}$$

and

$$T = \left(\frac{\mu m_p}{2k_B}\right) v_{circ}^2, \tag{16}$$

the same result as given by equation (13).

## 3. Results

### 3.1. Physical picture

To illustrate the effects of the UV background on the trajectories and temperature histories of fluid shells, we performed runs for collapse redshift $z_c = 2$ and two values of the circular velocity, $v_{circ} = 35\,\mathrm{km\,s^{-1}}$ ($M_f = 0.02 \times 10^{11} M_\odot$) and $v_{circ} = 128\,\mathrm{km\,s^{-1}}$ ($M_f = 10^{11} M_\odot$). In each case we ran the code once without a UV background ($J_{-21} = 0$) and once with a redshift-independent

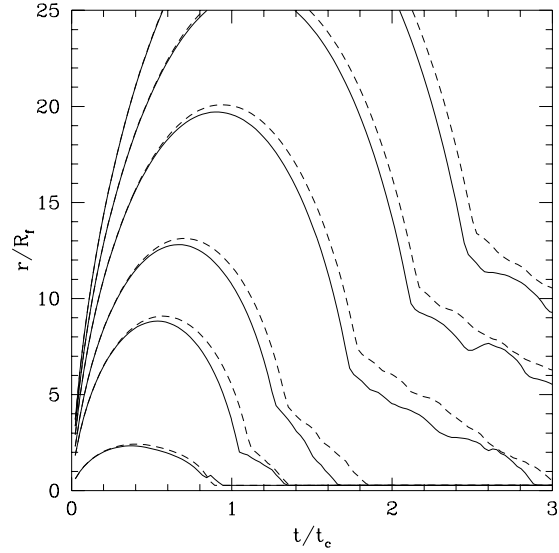

Fig. 3.— Fluid shell trajectories in high mass collapses ($M_f = 1 \times 10^{11} M_\odot$, $z_c = 2$, , $v_{circ} = 128\,\mathrm{km\,s^{-1}}$) with (dashed lines) and without (solid lines) a UV background. The time and radii are scaled with respect to the collapse time $t_c$ and the filter radius $R_f$.

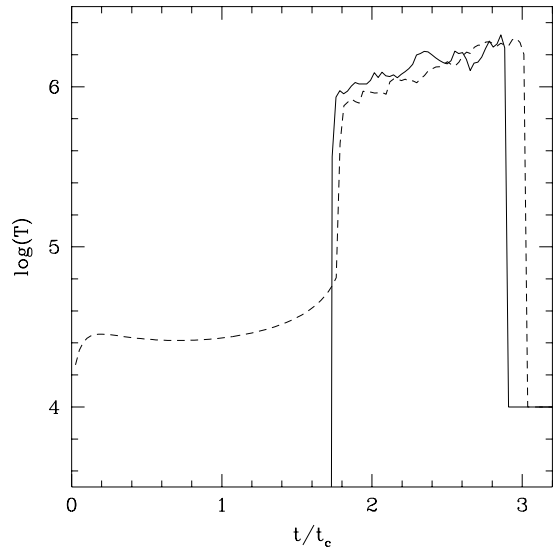

Fig. 4.— Time evolution of the temperature of a typical fluid shell in the high mass collapses of Fig. 3, with (dashed lines) and without (solid lines) a UV background.

UV background of amplitude $J_{-21} = 1$ and spectral slope $\alpha = 1$. The starting redshift for all simulations in this paper is $z_i = 36$.

Figure 3 shows the trajectories of six widely spaced fluid shells from the high mass runs. The time and radii are scaled with respect to the collapse time $t_c$ and the filter radius $R_f$. The solid lines represent trajectories obtained with no UV background, while the dashed lines are the trajectories for $J_{-21} = 1$. There is little difference between the solid and dashed trajectories. Figure 4 shows the time history of the temperature of a typical fluid shell in the absence (solid line) and presence (dashed line) of the UV background. When the UV background is present, the gas is preheated to a temperature of about $25,000$ K. Since this temperature is an order of magnitude smaller than the temperature $T \sim 6 \times 10^5$ K associated with a circular velocity of $128$ km s$^{-1}$, this preheating does not affect the collapse of the gas. Therefore, as illustrated in Figures 3 and 4, the UV background has essentially no effect on high mass collapses, as we speculated in §2.1 on the basis of the cooling curves. The physics underlying the trajectories and temperature histories in the no-ionization case is discussed in Paper I.

In Figure 5 we show fluid shell trajectories from the low mass collapses. Once again, the solid lines are trajectories for no UV background, and the dashed lines are trajectories for $J_{-21} = 1$, $\alpha = 1$. This time the trajectories are very different with and without the UV background. Figure 6 shows the temperature history of a typical fluid shell in the absence (solid line) and presence (dashed line) of the UV background. When the UV background is present, the gas is again preheated to a temperature of about $25,000$ K. However, this temperature is now of the same order of magnitude as the temperature $T \sim 45,000$ K associated with a circular velocity of $35$ km s$^{-1}$. Pressure support from the warm gas delays turnaround substantially and slows the subsequent collapse, greatly reducing the amount of baryonic mass that is able to collapse, cool, and sink to the center in a given amount of time. The dynamics of the collisionless dark matter shells (not shown) are almost identical with and without the UV background, since the baryons provide only a small fraction of the gravitating mass.

Figures 5 and 6 show that photoionization can dramatically alter the evolution of a low mass perturbation. The fact that these effects appear *before* a gas shell shock heats and begins to cool, indeed before the shell even turns around, strongly suggests that the UV background exerts its influence by heating the pre-collapse gas rather than by reducing cooling rates. The heavy dashed lines in Figures 5 and 6 confirm this view. They show trajectories and a temperature history from a run with $J_{-21} = 1$ in which we use photoionized abundances to compute cooling rates but set all *heating* terms to zero. The trajectories in this case are nearly identical to those obtained with no UV background at all. In this physically artificial scenario, cooling is possible at $T < 10^4$K because the photoionized gas emits recombination and Bremsstrahlung radiation and has no compensating heat input, so the gas cools well below $10^4$K and shells collapse slightly faster than they do when $J_{-21} = 0$.

Since the equilibrium temperature is lower for a softer UV spectrum, we expect the impact of the UV background to be smaller (see §2.1). The dotted lines in Figures 5 and 6 confirm this expectation. They show results for the same perturbation ($v_{circ} = 35\,\mathrm{km\,s^{-1}}$, $z_c = 2$) and UV background amplitude ($J_{-21} = 1$) but with spectral slope $\alpha = 5$ instead of $\alpha = 1$. Pressure support again delays turnaround and collapse, but the pre-heat temperature is lower and the resulting shell trajectories are intermediate between those obtained for no UV background and those obtained for $\alpha = 1$.

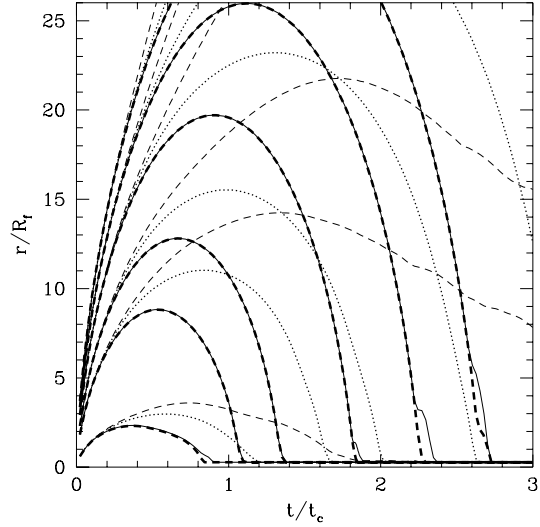

Fig. 5.— Fluid shell trajectories in a low mass collapse ($M_f = 0.02 \times 10^{11} M_\odot$, $z_c = 2$, $v_{circ} = 35 \, \mathrm{km \, s^{-1}}$). The solid lines are obtained in the absence of a UV background ($J_{-21} = 0$). The dashed lines are obtained with $J_{-21} = 1$ and $\alpha = 1$. The dotted lines are obtained with $J_{-21} = 1$ and $\alpha = 5$. The thick dashed lines are obtained with $J_{-21} = 1$ and $\alpha = 1$, but neglecting heating terms. The time and radii are scaled with respect to the collapse time $t_c$ and the filter radius $R_f$.

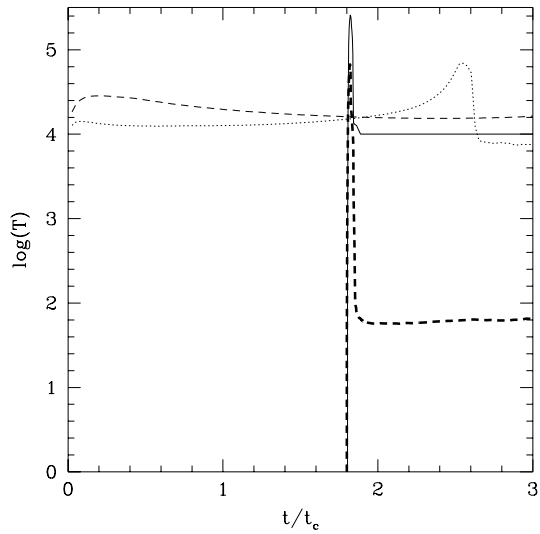

Fig. 6.— Temperature histories of a typical fluid shell in the collapses of Fig. 5, with no UV background (solid line), $J_{-21} = 1$ and $\alpha = 1$ (dashed line), $J_{-21}$ and $\alpha = 5$ (dotted line), and $J_{-21} = 1$, $\alpha = 1$ with heating terms neglected (thick dashed lines).

### 3.2.  Suppression of low mass galaxy formation

To study the implications of photoionization for galaxy formation, we have performed several series of collapse calculations. Within each series we vary the value of the filter mass $M_f$, and from one series to another we vary other parameters such as the collapse redshift and the amplitude, spectral slope, and time history of the UV background. In all of these calculations, we use $\Omega_d = 0.9$ and $\Omega_b = 0.1$. In a hierarchical model of structure formation, a dark matter halo typically merges with another dark halo of comparable size in a time of order its own collapse time $t_c$ (Lacey & Cole 1993). Gas that has not cooled and condensed by the time a merger occurs will typically be shock heated to the virial temperature of the new, merged halo. As our primary diagnostic, therefore, we show the mass $M_c$ of gas that cools and collapses to $r = 0$ by time $t = t_c$, $2t_c$, and $3t_c$ as a function of the circular velocity $v_{circ}$. Here, $t_c = t_0/(1 + z_c)^{3/2}$, where $t_0$ is the present age of the universe and $z_c$ is the collapse redshift as defined in §2.2. We scale $M_c$ to the mass $M_c(p = 0)$ of gas that would cool and collapse to $r = 0$ in the same time if radiative cooling were perfectly efficient (i.e., pressure equal to zero).

Figure 7 illustrates the effects of the UV background amplitude $J_{-21}$. It shows the ratio $M_c/M_c(p = 0)$ for runs with no ionizing background (crosses), with $J_{-21} = 1$ (circles), and with $J_{-21} = 0.3$ (asterisks). In all cases $z_c = 2$, and in the last two cases the spectral slope is $\alpha = 1$. As discussed already in §3.1, the photoionizing background inhibits the formation of low mass galaxies but does not change $M_c$ in high mass collapses. Furthermore, there is little difference between the results obtained with $J_{-21} = 0.3$ and those obtained with $J_{-21} = 1$. In both cases, we find that baryon collapse is completely suppressed below a cutoff circular velocity $v_{circ} \approx 30\,\mathrm{km\,s^{-1}}$ at $t = 3t_c$. At $t = 2t_c$, $M_c$ is reduced by about 50% for $v_{circ} \approx 50\,\mathrm{km\,s^{-1}}$.

To assess the effects of the shape of the UV spectrum, we compare runs with $\alpha = 1$, a hard spectrum that might be expected if the UV background is produced mostly by quasars, and $\alpha = 5$, a softer spectrum more like that expected from star-forming galaxies. In both cases we use an amplitude $J_{-21} = 1$ and a collapse redshift $z_c = 2$. Figure 8 plots the results for no UV background (crosses), $\alpha = 1$ (circles), and $\alpha = 5$ (asterisks). The UV background inhibits the formation of low mass galaxies regardless of the value of $\alpha$. However, the effect is greater for $\alpha = 1$, as speculated in §§2.1 and 3.1. The cutoff circular velocity at $t = 3t_c$ is $v_{circ} \approx 30\,\mathrm{km\,s^{-1}}$ if $\alpha = 1$ but $v_{circ} \approx 22\,\mathrm{km\,s^{-1}}$ if $\alpha = 5$. The collapsed mass at $2t_c$ is reduced by about 50% at $v_{circ} \approx 54\,\mathrm{km\,s^{-1}}$ for $\alpha = 1$ and at $v_{circ} \approx 41\,\mathrm{km\,s^{-1}}$ for $\alpha = 5$.

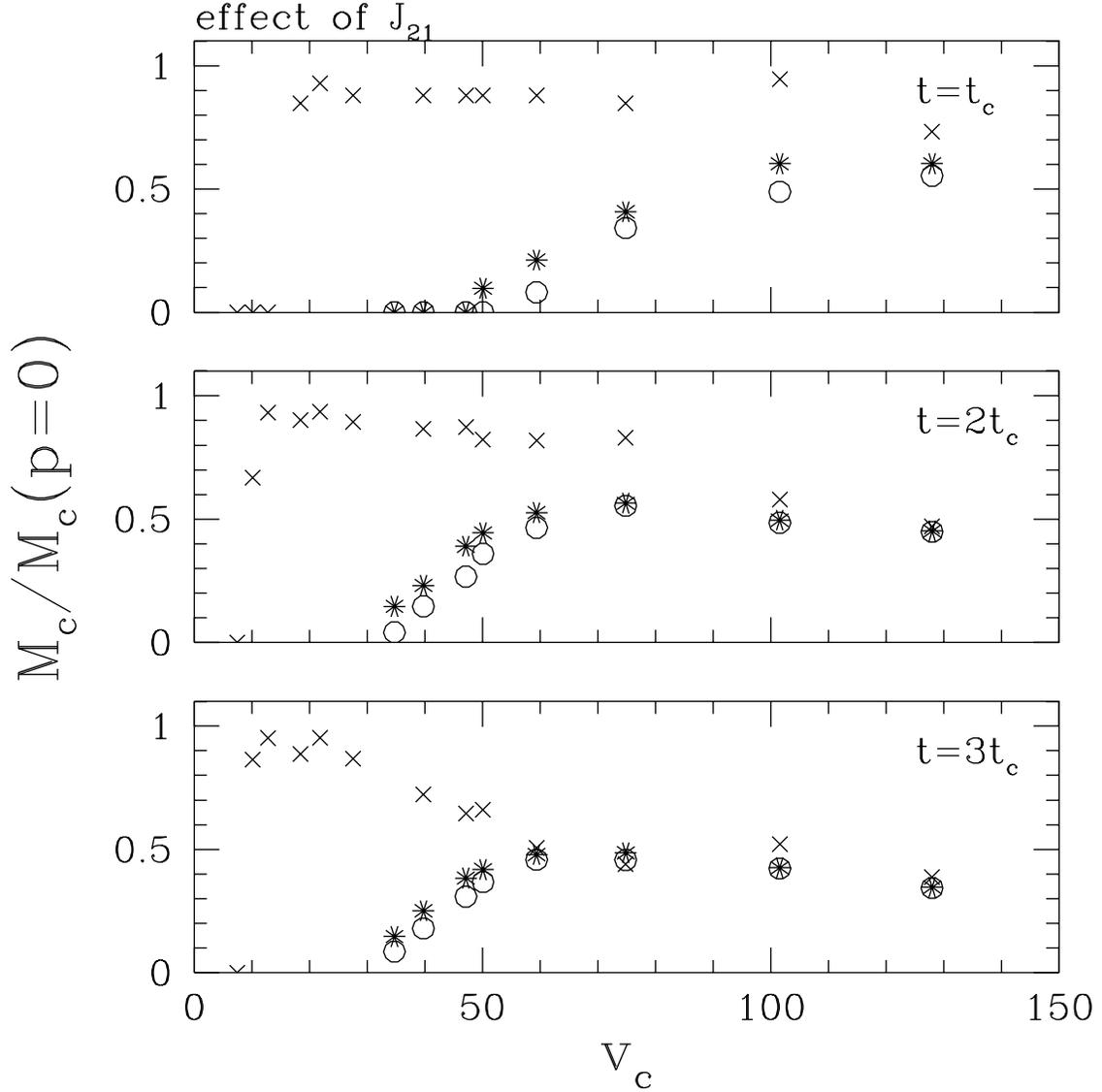

Fig. 7.— Mass $M_c$ of gas that cools and collapses by time $t = t_c, 2t_c, 3t_c$, as a function of the circular velocity $v_{circ}$ in km s$^{-1}$. Here $t_c = t_0/(1 + z_c)^{3/2}$, where $t_0$ is the present age of the universe, and we have taken $z_c = 2$. We have normalized $M_c$ to the mass $M_c(p = 0)$ of gas that would cool and collapse in the same time in the absence of pressure. The crosses are obtained in the absence of a UV background; the open circles are obtained in the presence of the photoionizing flux defined by equations (1) and (2), with $J_{-21} = 1$ and $\alpha = 1$; the asterisks are obtained for $J_{-21} = 0.3$ and $\alpha = 1$.

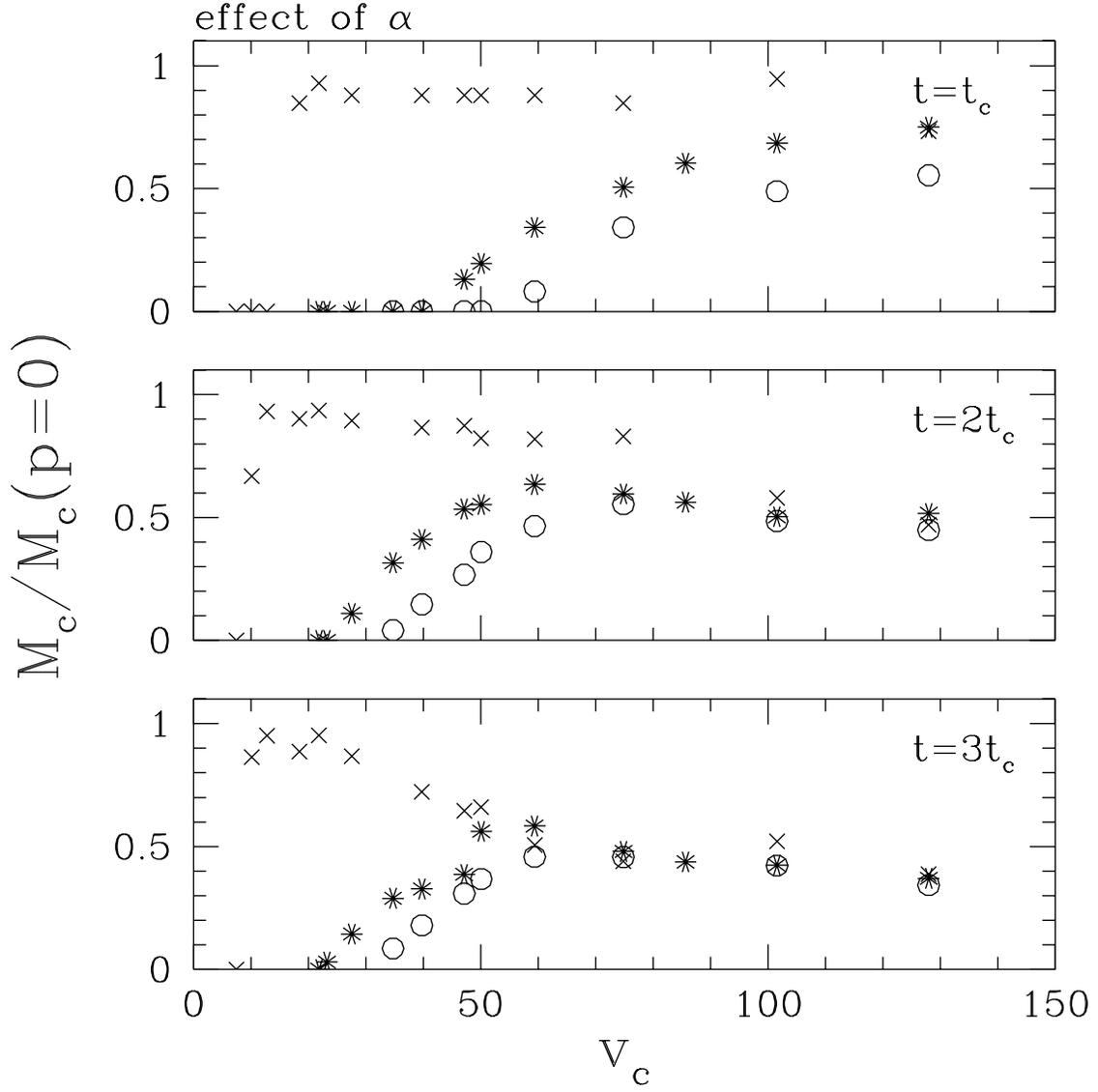

Fig. 8.— Like Fig. 7, but illustrating the dependence on the UV background spectral index $\alpha$. Crosses show results for no UV background; open circles show results for a UV background with $J_{-21} = 1$ and $\alpha = 1$, and asterisks for $J_{-21} = 1$ and $\alpha = 5$.

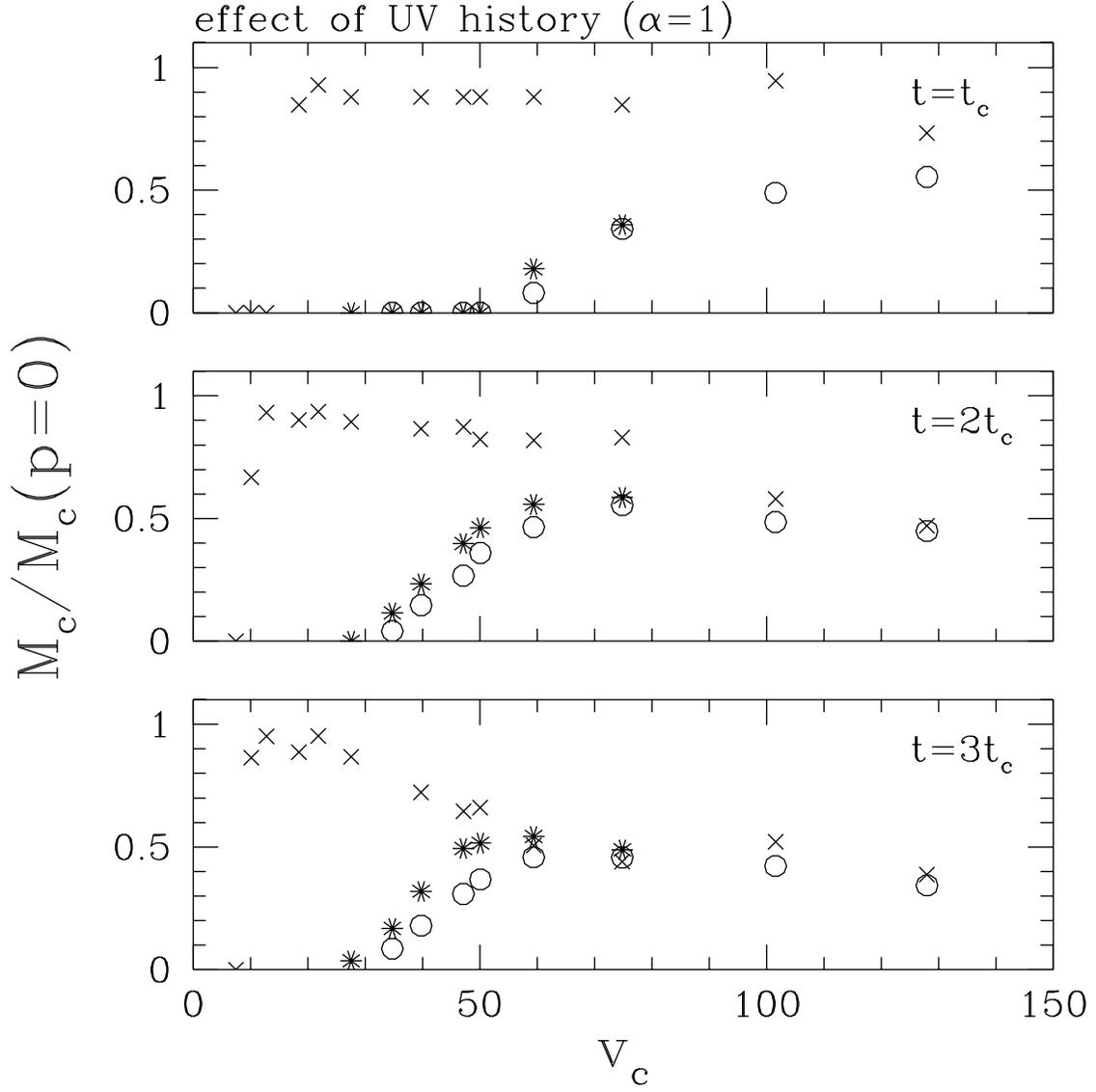

Fig. 9.— Like Fig. 7, but illustrating the dependence on the UV background time history. Crosses show results for no UV background, open circles show results for a constant photoionizing flux with $J_{-21} = 1$ and $\alpha = 1$, and asterisks show results obtained with the time-dependent UV flux defined by equation (17).

The assumption of a constant UV background intensity from $z = 36$ to $z = 0$ is unrealistic. To examine the effects of the UV background history, we compare the constant background ($J_{-21} = 1$) to a background that varies in roughly the way that might be expected if quasars are the dominant sources of the background:

$$J_{-21} = \begin{cases} 0 & \text{if } z > 6 \text{ ,} \\ \frac{4}{1+z} & \text{if } 3 < z < 6 \text{ ,} \\ 1 & \text{if } 2 < z < 3 \text{ ,} \\ \left(\frac{1+z}{3}\right)^3 & \text{if } z < 2 \text{ .} \end{cases} \tag{17}$$

In both cases we adopt a spectral slope $\alpha = 1$ and a collapse redshift $z_c = 2$. The results for the two models, plotted in Figure 9 together with the no background case, are not very different. Suppression by the UV background is slightly weaker when we use the history of equation (17) instead of a constant $J_{-21}$, since the gas in the former case is preheated by photoionization for a shorter time before turnaround. The effect is small, however, because the photoionizing background is turned on at a relatively high redshift, $z = 6$, and the amount of gas that turns around by this time is very small compared to the amount of gas that turns around by the collapse redshift $z_c = 2$. In both cases, the lower cutoff at $t = 3t_c$ is $v_{circ} \approx 30 \, \text{km s}^{-1}$, and the collapsed mass at $2t_c$ is reduced by about 50% at $v_{circ} \approx 50 \, \text{km s}^{-1}$.

In the case described above, the gas is assumed to be cold before the UV background turns on at $z = 6$, and the gas temperature thereafter evolves according to the heating and cooling rates predicted for ionization equilibrium. Miralda-Escudé & Rees (1994) point out that the process of reionization can heat gas more rapidly than this if it occurs fast enough. To examine the possible impact of such rapid heating, we compare the last series of runs to a series with the same UV background history (equation [17]) but with the gas temperature set instantaneously to an initial temperature $T_i = 50,000$K at $z = 6$. The results are shown in Figure 10. The value of the gas temperature at $z = 6$ has no effect on the amount of gas that collapses by $z = z_c$ and later. The equilibrium temperature at $z = 6$ is not much below $50,000$K, and the gas heats quickly to this temperature even from a cold start.

Figure 11 illustrates the effect of the collapse redshift, comparing runs with $z_c = 2$ and $z_c = 5$. In both cases $J_{-21} = 1$ and $\alpha = 1$. The effect of the background is slightly weaker for the higher collapse redshift. The cutoff at $t = 3t_c$ drops from $v_{circ} \approx 30 \, \text{km s}^{-1}$ for $z_c = 2$ to $v_{circ} \approx 25 \, \text{km s}^{-1}$ for $z_c = 5$, and the circular velocity for which $M_c$ at $t = 2t_c$ is reduced by 50% is $v_{circ} \approx 54 \, \text{km s}^{-1}$ for $z_c = 2$ and $v_{circ} \approx 46 \, \text{km s}^{-1}$ for $z_c = 5$. As predicted in §2.1, perturbations that collapse earlier are less affected by photoionization because their gas densities are higher and their equilibrium temperatures are correspondingly lower, but the difference is not striking. The difference would have appeared much larger if we had used mass $M_f$ instead of circular velocity $v_{circ}$ as the independent variable in Figure 11, but as argued in §2, circular velcity is the more natural parameter for this problem because of its direct connection to temperature.

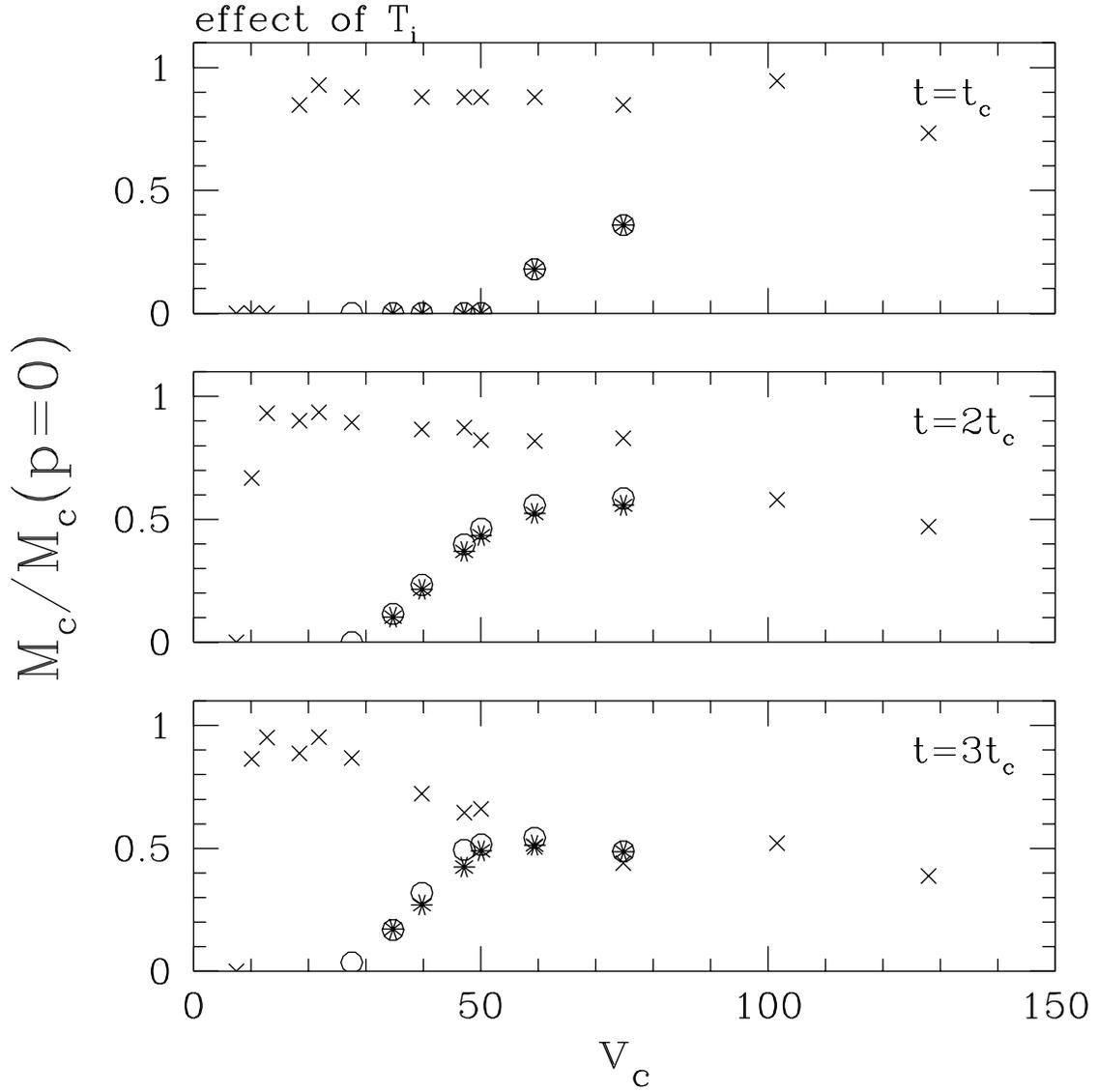

Fig. 10.— Like Fig. 7, but illustrating the effect of the gas temperature when the time-dependent UV background flux is turned on. Crosses show results for no UV background. Open circles and asterisks show results obtained with the time-dependent UV flux defined by equation (17), with an initial temperature at $z = 6$ of $T_i = 0$ and $T_i = 50,000$K, respectively.

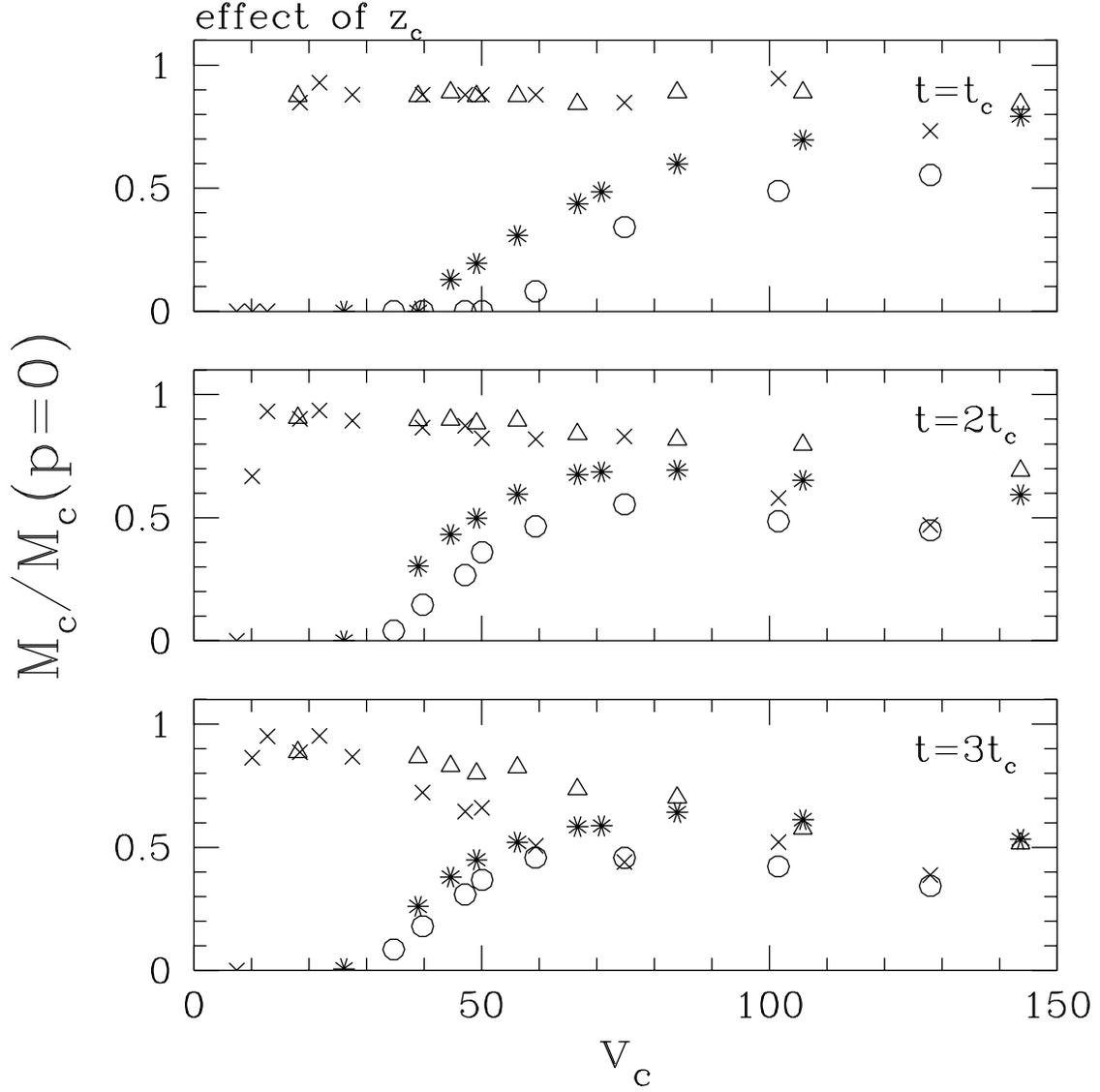

Fig. 11.— Like Fig. 7, but illustrating the dependence on the collapse redshift $z_c$. Crosses and triangles show results for no UV background with $z_c = 2$ and $z_c = 5$, respectively. Open circles ($z_c = 2$) and asterisks ($z_c = 5$) show results for a UV background with $J_{-21} = 1$ and $\alpha = 1$. In all cases $t_c = t_0/(1 + z_c)^{3/2}$, where $t_0$ is the present age of the universe, and $v_{circ}$ is defined by equation (10).

## 4. Discussion

Table 1 summarizes the results illustrated in Figures 7-11. For each series of runs, we list the circular velocity at which the cooled mass is reduced by 50% compared to its value in the absence of a UV background, and the cutoff circular velocity at which the baryon collapse is completely inhibited by the UV background, always at time $t = 2t_c$. The main result is a reduction of about 50% in the cooled mass at $v_{circ} \sim 50$ km s$^{-1}$, and complete suppression at $v_{circ} \sim 25 - 30$ km s$^{-1}$. As the table shows, this result is remarkably insensitive to the UV background parameters or collapse redshift. The most significant parameter is the spectral slope $\alpha$, which determines the equilibrium temperature of the gas (see Figure 1). Because of the robustness of this result, we expect that it is also insensitive to some of the parameters that we have not varied, such as $\Omega$ and the shape of the initial density perturbation.

| model | | | results at $t = 2t_c$ | |
|---|---|---|---|---|
| $z_c$ | $J_{-21}$ | $\alpha$ | $v_{circ}(50\%)$ | $v_{circ}$ (cutoff) |
| 2 | 1 | 1 | 55 km s$^{-1}$ | 34 km s$^{-1}$ |
| 2 | 0.3 | 1 | 49 km s$^{-1}$ | 30 km s$^{-1}$ |
| 2 | 1 | 5 | 41 km s$^{-1}$ | 24 km s$^{-1}$ |
| 5 | 1 | 1 | 46 km s$^{-1}$ | 27 km s$^{-1}$ |
| 2 | time-dependent $T_i = 0$ | 1 | 49 km s$^{-1}$ | 28 km s$^{-1}$ |
| 2 | time-dependent $T_i = 50,000$ K | 1 | 49 km s$^{-1}$ | 29 km s$^{-1}$ |

Table 1: Summary of the results at time $t = 2t_c$ for different values of the parameters: the collapse redshift $z_c$ and the amplitude $J_{-21}$, spectral slope $\alpha$, and time history of the UV background. We list the circular velocity $v_{circ}$ at which the cooled mass is reduced by 50% compared to its value in the absence of a UV background, and the cutoff circular velocity at which the baryon collapse is completely inhibited by the UV background.

Our study confirms Efstathiou's (1992) suggestion that photoionization can suppress the formation of low mass galaxies. The simulations yield quantitative estimates of the effect that are difficult to obtain in a robust way from Efstathiou's analytic approach. We are also able to clarify the physics of the effect: even though photoionization alters the cooling rates, it is clear from Figure 5 that it is the *heating* of the gas that matters. If the gas is able to overcome pressure support and collapse, it cools rapidly despite lowered cooling rates. Figure 5 also shows why the largest affected masses are somewhat higher than one might naively guess by equating the perturbation's virial temperature to the gas equilibrium temperature. Pressure support delays the turnaround of a gas shell, and the density at turnaround is lower than the density at virialization, so the characteristic dynamical temperature associated with the perturbation (which scales as $T \sim \sqrt{M/\rho^{1/3}}$) is somewhat higher.

Our results are consistent with the three-dimensional, SPH simulations of Weinberg, Hernquist & Katz (1995), who find no suppression of galaxy formation at $z = 2$ down to their resolution limit, which corresponds to a circular velocity $v_{circ} \approx 100 \,\mathrm{km\,s^{-1}}$. Our results imply that their minimum resolved mass is still substantially above the mass where photoionization influences galaxy formation. Still more reassuring, our results are consistent with the high-resolution SPH simulations of individual collapses by Quinn, Katz, & Efstathiou (1995). They find a drop in the cooled baryon mass of $\sim 30\%$ at $v_{circ} \sim 45 \,\mathrm{km\,s^{-1}}$, and complete suppression of baryon collapse at $v_{circ} \sim 25 \,\mathrm{km\,s^{-1}}$, values remarkably similar to those in Table 1. The two studies complement each other nicely. Quinn et al.'s results indicate that our conclusions are not sensitive to the idealizations of spherical symmetry and smooth initial conditions. Our results indicate that the suppression found by Quinn et al. is not sensitive to the numerical resolution (inevitably lower in 3-d simulations) or to their assumed parameters of the UV background. In particular, since the effects of the UV background depend mainly on the equilibrium temperature, the strong background intensity ($J_{-21} = 10$) assumed by Quinn et al. should not make much difference to the scale of suppression. Vedel et al. (1994) find that the internal structure of Milky Way-size galaxies can be affected by the presence of a UV background. Although they attribute this effect to the lower cooling rates, it is more likely a result of the reduction in small scale substructure during the collapse because of gas heating.

By suppressing collapse into low mass objects, photoionization can solve one problem that has faced hierarchical models of structure formation, a tendency to cool nearly all gas into small systems at very high redshifts. If first generation stars ionize the surrounding intergalactic medium, they will delay the formation of the next generation of objects until perturbations larger than the Jeans mass at $T \sim \mathrm{few} \times 10^4 \mathrm{K}$ are able to collapse. It is less clear that photoionization can solve the problems of hierarchical models in reproducing the observed galaxy luminosity function, since Press-Schechter (1974) type calculations indicate a rise above the estimated luminosity function already at $v_{circ} \sim 100 \,\mathrm{km\,s^{-1}}$ (White & Frenk 1991). However, photoionization will certainly help to reduce the discrepancy, and estimates of the slope of the local luminosity function may go up as low surface brightness galaxies are taken into account more accurately. On the theoretical side, a natural next step is to incorporate the results of collapse calculations like those presented here into Monte Carlo models of galaxy formation history (Kauffmann et al. 1993; Cole et al. 1994; Heyl et al. 1995). The inclusion of realistic photoionization and cooling calculations will solidify the physics underlying the Monte Carlo models, and the models will provide a larger cosmological framework within which to place the results of individual collapse simulations. The combination will allow better comparisons between the galaxy population predicted from theory and the galaxies in the local and high-redshift universe.

We thank Shaun Cole, Carlos Frenk, Neal Katz, Lars Hernquist, Jeremiah Ostriker, Tom Quinn, Martin Rees, and Simon White for helpful discussions. Many of these took place in the stimulating atmosphere of the Aspen Center for Physics. This research was supported by the Ambrose Monell Foundation (AT) and the W.M. Keck Foundation (DW) and by NSF grant

PHY92-45317 and NASA Astrophysical Theory Grants NAG5-2882 and NAG5-3085.